\documentclass[reprint]{JASAnew}

\usepackage{comment}

\newcommand{\ethan}[1]{\textcolor{black}{#1}}

\begin{document}

\title{Resolvent-based modeling of turbulent jet noise}
\author{Ethan Pickering}\thanks{Postdoctoral Associate}\email{epickeri@mit.edu}
\affiliation{Division of Engineering and Applied Science, California Institute of Technology, Pasadena, CA, 91125, USA}
\author{Aaron Towne}\thanks{Assistant Professor of Mechanical Engineering}
\affiliation{University of Michigan, Ann Arbor, MI, USA}
\author{Peter Jordan}	\thanks{CNRS Research Director}		
\affiliation{Institut Pprime, CNRS / Universit\'e de Poitiers /ENSMA, 86962 Futuroscope Chasseneuil, France}
\author{Tim Colonius} \thanks{Frank and Ora Lee Marble  Professor of Mechanical Engineering}
\affiliation{Division of Engineering and Applied Science, California Institute of Technology, Pasadena, CA, 91125, USA}


\date{\today} 

\begin{abstract}
    Resolvent analysis has demonstrated encouraging results for modeling coherent structures in jets when compared against their data-educed counterparts from high-fidelity large-eddy simulations (LES). We formulate resolvent analysis as an acoustic analogy that relates the near-field \ethan{resolvent} forcing to the near\ethan{- and far}-field pressure. We use an LES database of round, isothermal, Mach 0.9 and 1.5 jets to produce an ensemble of realizations for the acoustic field that we project onto a limited set of resolvent modes. In the near-field, we perform projections on a restricted acoustic output domain, $r/D = [5,6]$, while the far-field projections are performed on a Kirchhoff surface comprising a 100-diameter arc centered at the nozzle. This allows the LES realizations to be expressed in the resolvent basis via a data-deduced, low-rank, cross-spectral density matrix. We find that a single resolvent mode reconstructs the most energetic regions of the acoustic field across Strouhal numbers, $St = [0-1]$, and azimuthal wavenumbers, $m=[0,2]$. Finally, we present a simple function that results in a rank-1 resolvent model agreeing within 2dB of the peak noise for both jets.

\end{abstract}

\maketitle

\section{Introduction}

The goal of this work is to develop jet-noise models founded upon the physics of turbulent flows that are both low-rank and that provide insights into the mechanisms primarily responsible for noise generation. Resolvent analysis \citep{mckeon2010critical}, also known as input-output analysis \citep{jovanovic2020bypass}, provides a useful framework for achieving these goals. The central idea of the resolvent framework is similar to that of an acoustic analogy \cite{goldstein2003generalized,lighthill1952sound}, whereby a forcing term, related to the statistics of the hydrodynamic near-field turbulence, gives rise, through a linear operator, to the observed far-field sound.  The resolvent framework differs in two important ways.  First, the operator is decomposed into its singular components that represent the maximal amplification between the forcing and the output.  This permits the resulting acoustic field to be described \ethan{as} low rank, and thus limits the number of forcing statistics that must be modeled.  Secondly, the full linearized Navier-Stokes equations are used as the propagator, and we seek a modal basis that represents \textit{both} near and far-field coherent structures.  

Before recent advances in computational power, the idea of modeling both the hydrodynamic component along with the acoustics would have been seen as both unnecessary and computationally taxing. However, the ability to resolve both components of the flow is in fact a benefit. Starting with the experimental findings of \citet{mollo1967jet} and \citet{crow1971orderly}, it has become clear that coherent structures in the hydrodynamic near-field are directly responsible for far-field sound \citep{jordan2013wave}. These structures take the spatio-temporal form of wavepackets and have been found to be the dominant source for aft-angle sound \citep{jordan2013wave}, as well as partial contributors to sideline noise \citep{papamoschou2018wavepacket, jeun2018input}. These wavepackets may be linked to the early works of \citet{crighton1976stability} (and \citet{michalke1977instability}), who hypothesized that coherent structures could be described as linear instability modes of the mean flow via modal analysis. However, it has now become apparent that the correct representation of wavepackets is that of a highly-amplified response to turbulent fluctuations, which is directly found via the resolvent framework. 

Resolvent analysis uses the Singular Value Decomposition (SVD) to decompose the linear resolvent operator, identifying sets of orthogonal \textit{forcing/input} and \textit{response/output} modes, and ranking them in terms of the corresponding energetic gain between the forcing and response. This is particularly important as it allows our model to self-select the most relevant amplification mechanisms for noise generation. This allows for a natural truncation of the resolvent basis that produces a reduced-order model, or in other words, a reduced-rank acoustic analogy.

Several studies have applied resolvent analysis to develop low-rank jet models \citep{jeun2016input,cavalieri2019wave,lesshafft2018resolvent}.  The existence of relatively low-rank responses in round, turbulent jets was shown by \citet{SchmidtJFM2018}, with significant agreement between structures found through spectral proper orthogonal decomposition \citep{towne2018spectral} (SPOD) of a high-fidelity experimentally-verified large-eddy simulations (LES) of jets \citep{bres2017unstructured,bres2018importance}. Of particular relevance to this study are ``acoustic resolvent modes'' induced by performing resolvent analysis with an output domain defined over a region where fluctuations are purely acoustic. Through implementation of an acoustic output domain, resolvent analysis is able to filter out energetic, but acoustically irrelevant structures in the near-field. \citet{jeun2016input} performed such an analysis and found that for a Mach 1.5 jet, at Strouhal number $St = 0.33$ and azimuthal wavenumber $m=0$, the first resolvent mode reconstructs $57\%$ of the acoustic energy, but through inclusion of the next 23 resolvent modes the reconstruction improved to $70\%$ of the acoustic energy. This study looks to perform a similar analysis, in that we compute many acoustic resolvent modes and assess how well they reconstruct the acoustic energy. However, we also look to reduce the rank of the far-field significantly with the use of an eddy-viscosity model \citep{pickering2021optimal} and generalize the performance of the resolvent framework across frequencies $St = 0-1$, azimuthal wavenumbers $m=[0-2]$, and for two turbulent jets at Mach numbers of 0.9 and 1.5.

For a resolvent jet model to fully reconstruct flow statistics, and in this case those of the acoustic field, a resolvent-based model must incorporate sub-optimal modes \citep{SchmidtJFM2018} and correctly describe correlations (i.e. covariance) between modes inherent to turbulent flow \citep{towne2020resolvent}. These correlations are analogous to the concept of ``jittering'', used to describe temporal modulations of acoustic sources, that has been shown to be critical for accurately describing the acoustic field in turbulent jets \citep{cavalieri2011jittering}. In our approach, such temporal modulations, or jittering, may be represented through second-order statistics via the statistical representation of the resolvent operator \citep{towne2018spectral}
\begin{equation}
     \bm{S}_{yy} =  \bm{R}  \bm{S}_{ff} \bm{R}^*,
    \label{eqn:res_stat_intro}
\end{equation}
where $ \bm{S}_{yy}$ and $\bm{S}_{ff}$ are the cross-spectral density \ethan{(CSD)} tensors of the response and the forcing respectively and $\bm{R}$ is the resolvent operator. This equation shows that if the forcing CSD, describing spatial correlations, can be modeled \citep{towne2017statistical,zare2017colour},  then the resolvent operator identically reconstructs the flow statistics, $\bm{S}_{yy}$.  If the forcing  were spatially uncorrelated, $\bm{S}_{ff} = \bm{\Lambda}$, where $\bm{\Lambda}$ is a diagonal matrix, then the eigenvectors of $\bm{S}_{yy}$, which are the SPOD modes of the outputs, are aligned with the eigenvectors of $\bm{R} \bm{R}^*$ \citep{towne2018spectral}, or the response modes of the resolvent operator, $\bm{R}$.  However, the uncorrelated condition is rarely met, resulting in discrepancies between resolvent and SPOD modes that must be resolved through modeling $\bm{S}_{ff}$. 


One approach for modeling $\bm{S}_{ff}$ has been through the inclusion of a turbulence model within the resolvent operator. This approach has been implemented via an eddy-viscosity model in several flow configurations, from wall-bounded \cite{hwang2010amplification,morra2019relevance} to free shear flows \citep{pickering2021optimal}. The latter study, quantifying the effect on turbulent jet modeling, found that the use of an eddy-viscosity model (utilizing only quantities available from RANS models) significantly improved the agreement between SPOD and resolvent modes, thus reducing the effort required to model the effective $\bm{S}_{ff}$ by diminishing the magnitude of the off-diagonal terms. We utilize the same eddy-viscosity model in the present work to better model the acoustic field.

This paper explores an approach to describe the coupling between resolvent modes that is necessary for reconstructing the acoustic field with a minimal set of resolvent modes. The coupling provides directional and energetic variability in acoustic radiation inherently important for noise prediction \citep{cavalieri2011jittering}. Determination of the coupling between modes is performed by leveraging an ensemble of LES realizations which are projected on to a limited (i.e. low-rank) set  of acoustic resolvent modes. From these projections we attain a (drastically) reduced-order cross-spectral density between the retained modes--a Hermitian, frequency-dependent matrix of size $n \times n$, where $n$ denotes the number of retained modes, that accurately represents the acoustic field.

Organization of the manuscript is as follows. We first briefly describe the LES databases used, the main details pertaining to resolvent analysis, and present the statistical description of the resolvent framework for reconstructing the acoustic field and estimating the reduced order covariance matrix in \S~\ref{sec:methods_acoustic}. In \S~\ref{section:results} we present resolvent modes and LES reconstructions in the resolvent basis for one frequency-wavenumber pair for the Mach 1.5 jet before generalizing the approach to both jets over $St = [0,1]$ and $m=[0,2]$, and to both the near- and far-field acoustic regions. In the near-field section we compare the impact of including a RANS eddy-viscosity model to the resolvent operator and find it presents a significantly more efficient resolvent basis. We then present results for the far-field, along an arc at $100D$ from the nozzle, and show that reconstructions for both jets may be found using only the optimal resolvent mode. Finally, we conclude with a discussion on how the correct forcing coefficients may be estimated for a predictive jet noise model.

\section{Methods} \label{sec:methods_acoustic}
\subsection{Large Eddy Simulation database}
The LES database and resolvent analysis are fully described in \citet{SchmidtJFM2018} and \citet{towne2018spectral}.  Transonic (Mach 0.9) and supersonic (Mach 1.5) jets were computed using the flow solver ``Charles''; details on numerical methods, meshing, and subgrid-models can be found in \citet{BresAIAA2018} and \citet{bres2017unstructured} along with validation cases conducted at PPRIME Institute, Poitiers, France for the Mach 0.9 jet \citep{bres2018importance}. The Mach 0.9 and 1.5 jets have Reynolds numbers of $Re_j = \rho_j U_j D / \mu_j =1.01 \times 10^6$ and $Re_j = 1.76 \times 10^6$, respectively, where subscript $j$ gives the value at the center of the jet, $\rho$ is density, $\mu$ is viscosity, and $M_j$ is the Mach number $M_j = U_j/c_j$, with $c_j$ as the speed of sound at the nozzle centerline.

Throughout the manuscript, variables are non-dimensionalized by the mean jet velocity $U_j$, jet diameter $D$, and pressure $\rho_j U_j^2$, with the resulting equation of state $p = \frac{\rho T}{\gamma M_j^2}$, with $T$ denoting temperature and $\gamma$ the ratio of specific heats. Frequencies are reported in Strouhal number, $St = f D / U_j$, where $f$ is the frequency \ethan{in Hertz}. The database consists of 10,000 snapshots separated by $\Delta t c_{\infty}/D = 0.2$ and 0.1 for the $M_j = 0.9$ and $M_j = 1.5$ jets, respectively, with $c_\infty$ as the ambient speed of sound, and interpolated onto a structured cylindrical grid $x,r,\theta \in [0,30] \times [0,6] \times [0, 2\pi]$, where $x$, $r$, $\theta$ are streamwise, radial, and azimuthal coordinates, respectively. Variables are reported by the vector
\begin{align}
\bm{q} = [\rho, u_x, u_r, u_\theta, T]^T,    
\end{align}
where $u_x$, $u_r$, $u_\theta$ are the three cylindrical velocity components.

To generate an ensemble of flow realizations for computing statistical averages, the LES database of 10,000 snapshots is segmented into bins of 256 snapshots, with an overlap of 75\%, and under the implementation of a Hamming window, resulting in 153 realizations of the flow. Each realization is then decomposed in the azimuthal direction and in time. The temporal decomposition provides a resolution of $St = 0.026$ and  $St = 0.0217$ the $M_j = 1.5$ and $M_j= 0.9$ jets, respectively, and the azimuthal decomposition is valid up to $m=68$; however, the acoustically relevant azimuthal wavenumbers are much smaller \citep{juve1979filtered} and only azimuthal wavenumbers $m=[0-2]$ are considered in this paper.  

Considering the LES database only extends to $r/D = 6$, we implement a Kirchhoff surface (\ethan{as described in \cite{freund2001noise}}), to the azimuthally and temporally transformed realizations of the flow. In doing so, we create an ensemble of far-field realizations located along an arc, with angle $\phi$, of $100D$ from the nozzle at each frequency and azimuthal wavenumber. As done in  \citet{bres2017unstructured}, and associated experiments \citep{Schlinker2008,Schlinker2009}, we specifically compute the acoustics for the aft-angle sound from $\phi = 100 - 160$ and find our acoustic far-field is in close agreement (within 2dB) with the far-field of the LES calculation. 


\subsection{Resolvent analysis}\label{sec:methods_resolvent}
For the round, statistically-stationary, turbulent jets considered in this manuscript, the compressible Navier-Stokes, energy, and continuity equations are linearized via a standard Reynolds decomposition  \ethan{(see \citet{pickering2021optimal} Appendix B for a detailed description of the governing equations in cylindrical coordinates)} and Fourier transformed both in time and azimuthally to the compact expression
\begin{equation}
(i\omega\textbf{I} - \textbf{A}_m) \bm{q}_{m, \omega} = \bm{L}_{m,\omega} \bm{q}_{m, \omega} = \bm{n}_{m, \omega},
\label{LNS_eq_acoustic}
\end{equation}
where \ethan{$\omega = 2 \pi St$ is the frequency,  $m$ is the azimuthal wavenumber, $\mathbf{I}$ is the identity matrix, $\mathbf{A}_{m}$ is the frequency independent linear operator,  $\bm{L}_{m,\omega}$ is the total forward linear operator, $\bm{q}_{m, \omega}$ is the response in each variable, and $\bm{n}_{m, \omega}$} constitutes the nonlinear forcing. Mean-flow quantities used in the operator are derived from a RANS model, fitted closely to the LES mean flow. Although the mean flows are similar, the computation of a RANS model, using the standard $\kappa-\epsilon$ closure equations, also provides an eddy-viscosity field \ethan{that may be included in the resolvent operator. This is done following results of \citet{pickering2021optimal} that presented} substantially improved agreement between SPOD and resolvent modes with the inclusion of an eddy-viscosity model. \ethan{The eddy-viscosity used here is computed} as $\mu_T = c C_{\mu} k^2/\epsilon$, where $c$ and $C_{\mu}$ are scaling constants ($c=0.2$, $C_{\mu} = 0.0623$ for the $M_j = 0.9$ and $C_{\mu} = 0.0554$ for $M_j = 1.5$ jet), $k$ is the turbulent kinetic energy field, and $\epsilon$ is the turbulent dissipation field. \ethan{We stress that the values of $C_{\mu}$ used here were selected to reproduce the LES mean flow and to allow for a demonstration of the ability of quantities available in RANS to be used to compute accurate resolvent modes. These values should not be considered general or as recommended values of $C_{\mu}$ for general purposes.} 

Continuing with the derivation of the resolvent/input-output operator, we rewrite equation \eqref{LNS_eq_acoustic} by moving $\bm{L}_{m,\omega}$ to the right-hand side to give
\begin{equation}
\bm{q}_{m, \omega}  = \bm{L}_{m,\omega}^{-1} \bm{n}_{m, \omega} = \bm{R}_{m,\omega} \bm{n}_{m, \omega},
\label{eqn:resolvent_acoustic}
\end{equation}
where $\bm{R}_{m,\omega} = \bm{L}_{m,\omega}^{-1}$ is the standard resolvent operator. To then specify particular domains for both the response and forcing, we \ethan{replace $\bm{n}_{m,\omega}$ with $\bm{B} \bm{f}_{m, \omega}$, where $\bm{f}_{m, \omega}$ represents an (unrestricted) nonlinear forcing, to the above} as
\begin{equation}
    \bm{q}_{m, \omega}  = \bm{R}_{m,\omega}  \bm{B} \bm{f}_{m, \omega}
    \label{eqn:qRBF}
\end{equation}
and define the output variable
\begin{equation}
     \bm{y}_{m, \omega}  =  \bm{C} \bm{q}_{m, \omega},
     \label{eqn:yCq}
\end{equation}
where $\bm{B}$ and $\bm{C}$ are input and output matrices. \ethan{The latter matrix, $\mathbf{C}$, is used to isolate the acoustics in the near-field, or propagate fluctuations to the far-field. Each of these cases are detailed in Appendix \ref{app:C}.} Inserting equation \eqref{eqn:qRBF} into equation \eqref{eqn:yCq} gives the input-output relationship,
\begin{equation}
     \bm{y}_{m, \omega}  =  \bm{C} \bm{R}_{m,\omega}  \bm{B} \bm{f}_{m, \omega} = \bm{H}_{m,\omega} \bm{f}_{m, \omega},
     \label{eqn:yHq}
\end{equation}
where $\bm{H}_{m,\omega} = \bm{C} \bm{R}_{m,\omega}  \bm{B} $ is the resolvent input-output operator from $\bm{f}_{m, \omega}$ to $\bm{y}_{m, \omega}$. Then by introducing the compressible energy norm of \citet{chu1965energy},
\begin{align}
    &\langle \bm{q}_1, \bm{q}_2 \rangle_E \nonumber \\ & = \int_{\Omega}  \bm{q}_1^* \text{diag} \bigg( \frac{\bar{T}}{\gamma \bar{\rho} M^2}, \bar{\rho},  \bar{\rho}, \bar{\rho}, \frac{\bar{\rho}}{\gamma (\gamma - 1) \bar{T} M^2} \bigg) \bm{q}_2 \text{d}\Omega \nonumber \\ &= \bm{q}^*_1 \bm{W} \bm{q}_2,
\end{align}
(where \ethan{$\Omega$ is the domain volume and} superscript $*$ denotes the complex conjugate transpose) via the matrix $\bm{W}$ to the forcing and response ($\bm{W}_f = \bm{W}_y = \bm{W}$) the weighted resolvent input-output operator
\begin{equation}
\hat{\bm{H}}_{m, \omega}  = \bm{W}_y^{1/2} \bm{H}_{m,\omega} \bm{W}_f^{-1/2}
\end{equation}
is obtained. Resolvent modes may then be found by taking the singular value decomposition of the weighted resolvent input-output operator giving
\begin{equation}
    \hat{\bm{H}}_{m, \omega}  = \hat{\bm{U}}_{m,\omega} \bm{\Sigma}_{m,\omega} \hat{\bm{V}}_{m, \omega}^*,
\end{equation}
where the optimal response and forcing modes are contained in the columns of $\bm{U}_{m,\omega}  = \bm{W}_y^{-1/2} \hat{\bm{U}}_{m,\omega}$, with $\bm{U}_{m, \omega} = [\bm{u}_{m, \omega}^1, \bm{u}_{m, \omega}^2, ... , \bm{u}_{m, \omega}^N]$, $\bm{V}_{m,\omega}  = \bm{W}_f^{-1/2} \hat{\bm{V}}_{m,\omega}$, $\bm{V}_{m, \omega} = [\bm{v}_{m, \omega}^1, \bm{v}_{m, \omega}^2, ... , \bm{v}_{m, \omega}^N]$, and $\bm{\Sigma}_{m,\omega} = \text{diag}(\sigma_{m, \omega}^1, \sigma_{m, \omega}^2, ... , \sigma_{m, \omega}^N)$ are the optimal gains \cite{towne2018spectral}. The unweighted resolvent input-output operator may then be recovered as
\begin{equation}
    \bm{H}_{m, \omega}  = \bm{U}_{m,\omega} \bm{\Sigma}_{m,\omega} \bm{V}_{m, \omega}^* \bm{W}_f.
\end{equation}

\subsection{Statistics}
The statistics we are interested in are contained within the cross-spectral density tensor, which may be found for the desired output space by multiplying the resolvent equation by its complex conjugate transpose and taking the expectation \citep{towne2018spectral}
\begin{equation}
    \langle \bm{y}_{m, \omega} \bm{y}_{m, \omega}^* \rangle = \langle \bm{H}_{m, \omega} \bm{f}_{m, \omega} \bm{f}_{m, \omega}^* \bm{H}_{m, \omega}^*\rangle,
\end{equation}
giving 
\begin{equation}
     \bm{S}_{yy,m, \omega} =  \bm{H}_{m, \omega}  \bm{S}_{ff,m, \omega} \bm{H}_{m, \omega}^*,
    \label{eqn:res_stat}
\end{equation}
where $ \bm{S}_{yy,m, \omega}$ and $ \bm{S}_{ff,m, \omega}$ are the CSD tensors of the response and the forcing, respectively. For brevity, we drop the subscripts $m$ and $\omega$ and note that all CSD tensors and resolvent matrices must be defined for specific $m$ and $\omega$ pairs in the remainder of the manuscript.

As mentioned in the introduction, this representation shows that if the forcing CSD tensor is known, then the resolvent operator reconstructs the response statistics. However, the forcing CSD is generally unknown. There are at least two potential avenues for modeling it.  The first is to directly model $\bm{S}_{ff}$.  To aid in such modeling efforts, $\bm{S}_{ff}$ may be computed directly from full LES data \citep{towne2017statistical}, or estimated from limited flow statistics \cite{towne2020resolvent}.  A second approach is to modify the resolvent operator by supplementing the governing linearized equations with an appropriately linearized turbulence model.  In \citet{pickering2021optimal}, an eddy-viscosity model was considered and LES data was used to determine an optimal eddy viscosity field that would align, insofar as possible, the modes of $\bm{S}_{qq}$ (i.e. the full response statistics) with those of $\bm{R} \bm{R}^*$ (identical to $\bm{H} \bm{H}^*$ when $\bm{C} = \bm{B} = \bm{I}$). They found this to substantially reduce the magnitude of the off-diagonal terms of $\bm{S}_{ff}$, at least as the near-field coherent structures were concerned, consequently simplifying the number of terms that must be modeled.

In this study, we combine both modeling approaches. We first utilize the eddy-viscosity approximation of \citet{pickering2021optimal} and then estimate a low-order approximation of the forcing CSD for the acoustic field. To do the latter, we return to equation \eqref{eqn:res_stat} and expand the resolvent input-output operator through its singular value decomposition,
\begin{equation}
         \bm{S}_{yy} =  \bm{U} \bm{\Sigma}  \bm{V}^* \bm{W}_f \bm{S}_{ff}  \bm{W}_f \bm{V}  \bm{\Sigma}  \bm{U}^*
\end{equation}
and define a covariance matrix $ \bm{S}_{\beta \beta} =   \bm{V}^* \bm{W}_f  \bm{S}_{ff} \bm{W}_f \bm{V}$, where $\bm{\beta} =  \bm{V}^* \bm{W}_f \bm{f}$ is the projection of the forcing upon the resolvent input modes. This gives  
\begin{equation}
         \bm{S}_{yy} =  \bm{U} \bm{\Sigma}  \bm{S}_{\beta \beta}  \bm{\Sigma} \bm{U}^*,
\end{equation}
which can be rearranged to solve for the covariance matrix,
\begin{equation}
         \bm{S}_{\beta \beta} =  \bm{\Sigma}^{-1} \bm{U}^*  \bm{S}_{yy}  \bm{U} \bm{\Sigma}^{-1}.
\end{equation}
In its current state, the covariance matrix is exact, maintaining a full size of the system with approximately $10^{11}$ degrees of freedom (i.e. $\bm{S}_{\beta \beta} \in  \mathcal{C}^{5 N_x N_r \times 5 N_x N_r}$).  To obtain a low-rank model of $ \bm{S}_{\beta \beta}$ from the LES data, we compute $\bm{S}_{\beta \beta}$ with a truncated set of $n$ resolvent modes, $\tilde{\bm{U}} \in \mathcal{C}^{5 N_x N_r \times n}$, as,
\begin{equation}
    \tilde{\bm{S}}_{\beta \beta} =  \tilde{\bm{\Sigma}}^{-1} \tilde{\bm{U}}^*  \bm{S}_{yy}   \tilde{\bm{U}} \tilde{\bm{\Sigma}}^{-1}.
\end{equation}
This reduces the size of the covariance matrix to $n \times n$, drastically reducing the number of degrees of freedom to $O(10^0-10^1)$.

With $\tilde{\bm{S}}_{\beta \beta}$, we may ask several questions: How well does $\tilde{\bm{S}}_{\beta \beta}$ reconstruct $\bm{S}_{yy}$ \ethan{in the truncated resolvent basis, where the reconstructed CSD is computed as
\begin{equation}
    \tilde{\bm{S}}_{yy} = \tilde{\bm{U}}_y \tilde{\bm{\Sigma}} \tilde{\bm{S}}_{\beta \beta} \tilde{\bm{\Sigma}}  \tilde{\bm{U}}_y^*.
\end{equation}
M}ay $ \tilde{\bm{S}}_{\beta \beta}$ be further reduced (e.g. neglect off-diagonal terms) and can $ \tilde{\bm{S}}_{\beta \beta}$ be modeled? \ethan{These questions are addressed in the following sections.} 

\section{Results} \label{section:results}

\begin{figure*}
\begin{center}
	\includegraphics[width=1\textwidth,trim={0cm 0cm 0cm 0.15cm},clip]{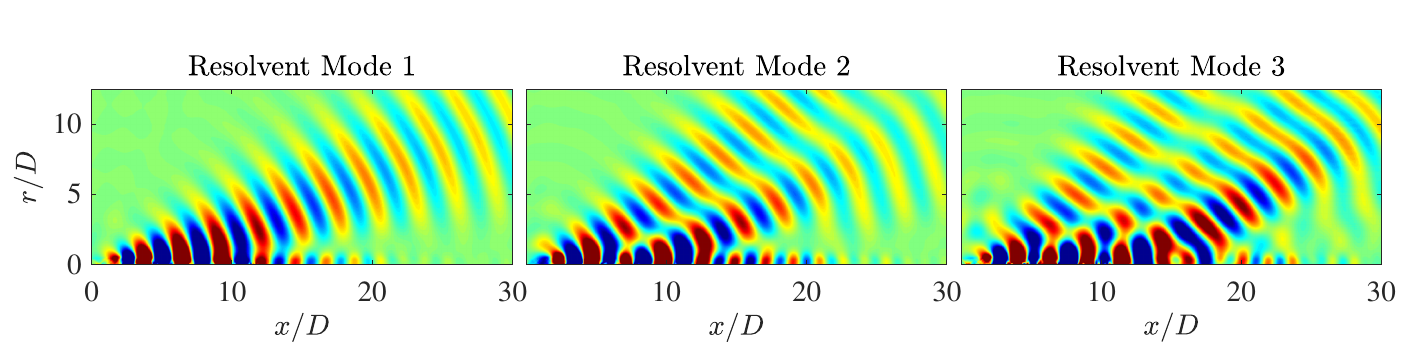}
\end{center} \vspace{-0.85cm}
\caption{The first three resolvent modes of fluctuating pressure, $q_{p'}$. Red and blue contours vary from $\pm$ $20\%$ of the maximum fluctuating pressure of each mode, $\pm 0.2 ||q_{p'}||_{\infty}$.  $M_j=1.5$, $St=0.26$, $m=0$.}
\label{fig:ResolventModes}
\end{figure*}

\begin{figure*}
\begin{center}
	\includegraphics[width=1\textwidth,trim={0cm 0cm 0cm 0cm},clip]{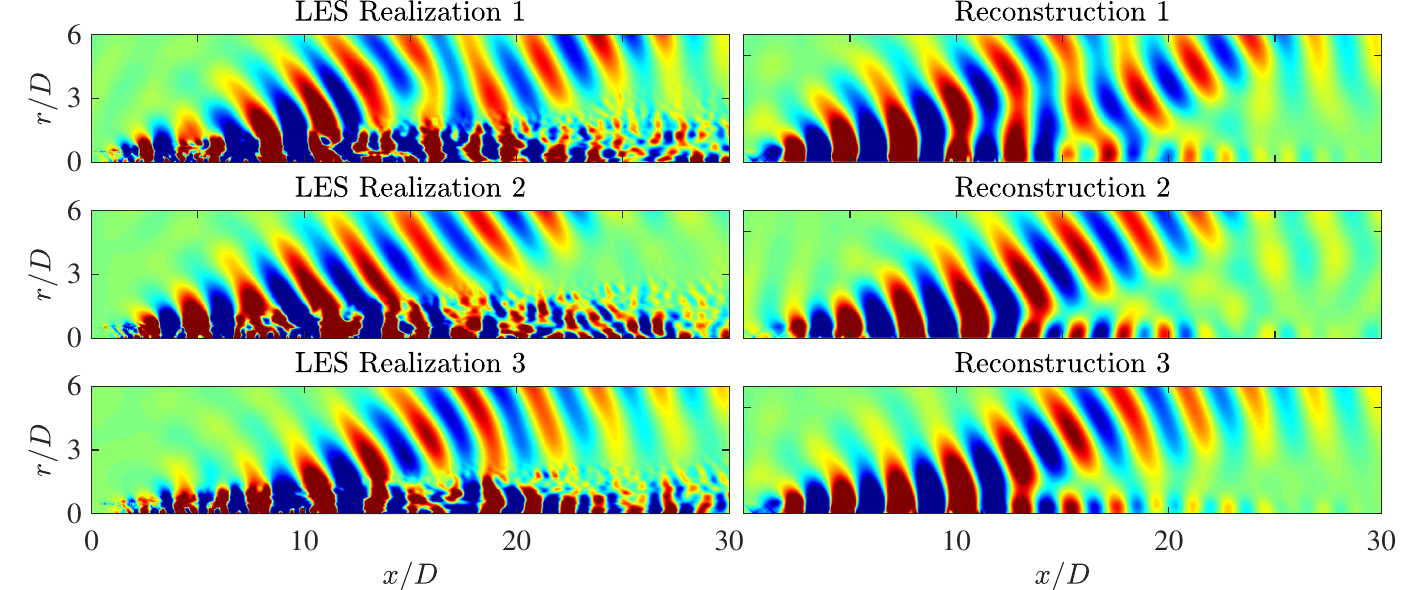}
\end{center} 
	\caption{Three LES realizations (left) and their associated three-mode resolvent reconstructions (right) of the pressure field at  $M_j=1.5$, $St=0.26$, $m=0$.  Red and blue contours vary from $\pm$ $20\%$ of the maximum fluctuating pressure of each LES realization, $\pm 0.2 ||q_{p'}||_{\infty}$.}
	\label{fig:LES_Realization_Recontruction}
\end{figure*}

\subsection{Near acoustic field} \label{sec:near-field}

We begin by providing detailed results for a single frequency and azimuthal wavenumber pair of the $M_j= 1.5$ jet using the RANS eddy-viscosity resolvent operator \ethan{and a near-field acoustic $\mathbf{C}$ output matrix ($r/D = [5,20]$ in the fluctuating pressure field, details provided in \ref{app:near-field})}. Figure \ref{fig:ResolventModes} presents the first three resolvent modes \ethan{for $M_j = 1.5$, $St = 0.26$, and $m=0$} computed with the restricted acoustic output domain and then recast in the full domain by {setting $\mathbf{C} = \mathbf{I}$ and calculating} $\bm{U}_{q} =  \bm{L}^{-1}  \bm{V}_{y}$. The associated gain of these modes, normalized by the first resolvent gain, are $[1, 0.17, 0.15]$ (and slowly decreasing with higher modes), indicating the first resolvent mode has at least six times the amplification as the following resolvent modes.

The resolvent response modes show a particular pattern of acoustic beams. For the first mode there is a single, energetic beam, propagating at a shallow angle to the jet axis. The first suboptimal mode consists of two beams, similar to what was found by \citet{jeun2016input}. This pattern is shown by the next suboptimal mode, with three beams located at the perimeter of the first suboptimal. Although not shown, this behavior continues for further suboptimal modes. 

Figure \ref{fig:LES_Realization_Recontruction} compares three specific \ethan{(but randomly chosen)} realizations of the $m=0$, $St=0.26$ field from the LES, $\bm{q}$, to the three-mode reconstructions of these fields found by projection. \ethan{The reconstructions are found by} $\tilde{\bm{q}} = \bm{\tilde{U}}_q \tilde{\bm{\alpha}}$, where $\tilde{\bm{\alpha}} =  \bm{\tilde{U}}_z^{+*} \bm{W}_z \bm{z}$ \ethan{and $\bm{z}$ is an acoustic subset of the LES domain ($r/D = [5,6]$ of the pressure field) and $\bm{\tilde{U}}_z^{+*}$ is the psuedoinverse projecting the LES domain $\bm{z}$ and resolvent output domain $\bm{y}$. Further details for using the psuedoinverse to project resolvent modes on other spaces is provided in Appendix \ref{app:non_orthogonal}.  From figure \ref{fig:LES_Realization_Recontruction} we see that the} three resolvent modes are able to accurately reconstruct the different radiation patterns evident in the LES realizations. Clearly there is constructive and destructive reinforcement amongst the three resolvent modes in order to produce the LES realizations.

\begin{figure*}
\begin{center}
\includegraphics[width=1\textwidth,trim={0cm 0cm 0cm 0cm},clip]{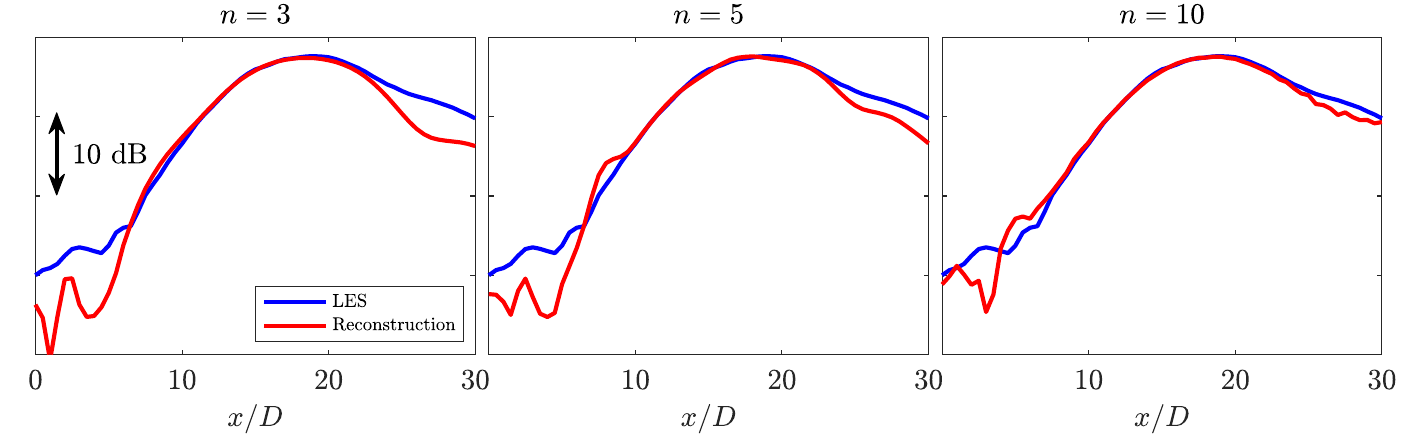}
\end{center}
\caption{Comparison of pressure PSD values by dB at $r/D = 6$ for the LES ensemble and reconstructions in the resolvent basis using 3, 5, and 10 resolvent modes.}
\label{fig:Reconstruction_Acoustics}
\end{figure*}
\ethan{
\begin{figure*}
\centering
\includegraphics[width=1\textwidth]{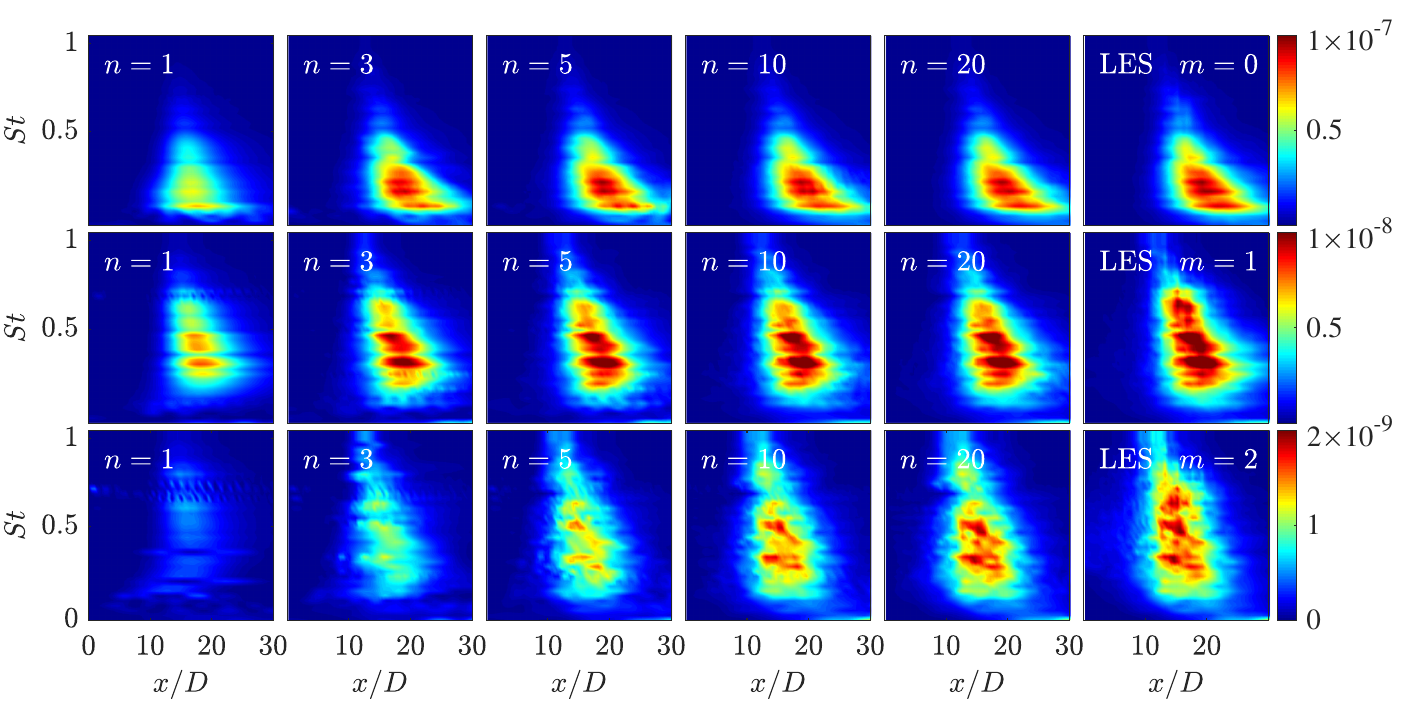}
\caption{PSD of resolvent reconstructions of the $M_j = 1.5$ jet with a RANS eddy-viscosity model at radial surface $r/D=6$ from $St = [0,1]$ and $x/D = [0,30]$ for three azimuthal wavenumbers, $m=[0, 2]$, from top to bottom and using $n=1,3,5,10,20$ modes from left to right. The right most column presents the LES values and the contour levels associated with each row.}
\label{fig:M15_Recon_New}
\end{figure*}}

\ethan{For a more quantitative assessment of the ability of the resolvent modes to reconstruct the acoustic field, we compute and compare the power spectral density (PSD) of the acoustic field, which is located in the diagonal terms of $\bm{S}_{yy}$. Specifically, we report the difference between the two PSD in decibels, 
\begin{equation}
    \Delta \text{dB} = 10\text{log} \bigg( \text{diag}(\tilde{\bm{S}}_{yy} - \bm{S}_{yy})\bigg),
\end{equation}
at $r/D = 6$ in figure \ref{fig:Reconstruction_Acoustics}.} This is again performed for $St=0.26$, $m=0$, but is now averaged over all $k = 153$ realizations. In addition to the three resolvent mode set, results are also shown for 5 and 10 mode sets. With just three modes we see that the peak directivity is well captured, with minor improvements (and diminishing returns) in the off-peak directivity with increasing numbers of modes.

\begin{figure}
\vspace{-0.25cm} \hspace{0.65cm} $M_j = 0.9$ \hspace{2.35cm} $M_j = 1.5$ \\ 
	\includegraphics[width=0.495\textwidth,trim={0cm 0cm 0cm 0cm},clip]{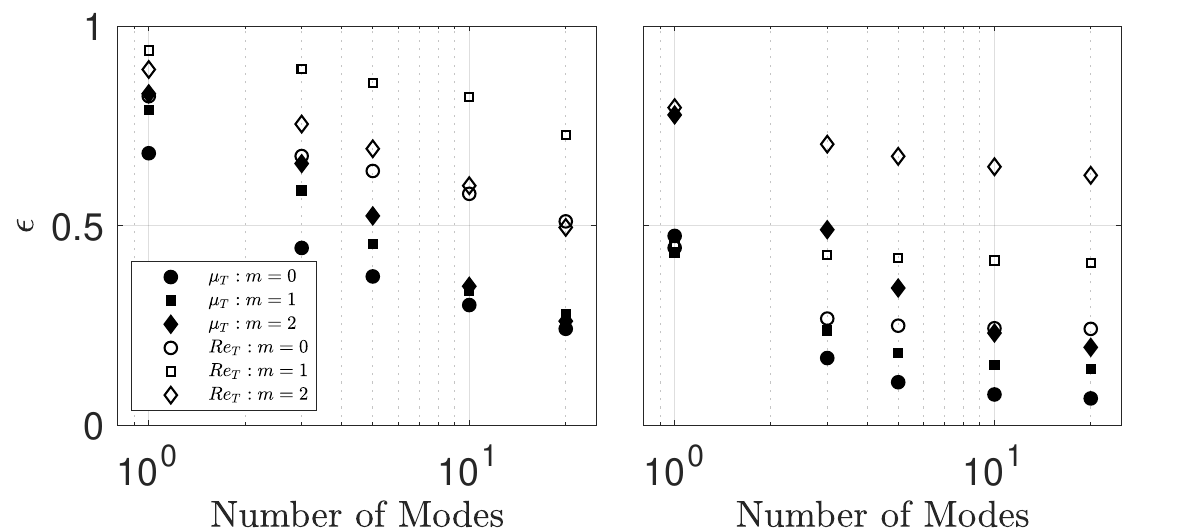}
	\vspace{-0.5cm}
\caption{\ethan{Error between the LES PSD and the reconstructed PSD by number of resolvent modes retained (i.e. $n = [1,3,5,10,20]$). Filled symbols indicate modes using the RANS eddy-viscosity model, while hollow symbols give those using a turbulent Reynolds number, $Re_T = 3 \times 10^4$.}}
\label{fig:NearFieldEnergy}
\end{figure}

We now extend our comparison to Strouhal numbers ranging from 0 to 1 and azimuthal wavenumbers 0-2 and assess the overall ability of the truncated resolvent basis to reconstruct the acoustic field. Figure \ref{fig:M15_Recon_New} compares the PSD from the LES to its $n$-rank resolvent-basis reconstructions with $n=1$, 3, 5, 10, and 20.  The rank-1 model results are similar to those of Sinha et al. \citep{sinha2014wavepacket}, who used parabolized stability equations and projected onto the first SPOD-mode at each $St-m$ pair.  However, we show here that once additional modes are included, the reconstructions are substantially improved: the 20-mode model shows close agreement with the LES for all frequencies and azimuthal modes, while even the 3-mode model is qualitatively accurate for $m=0$ and $m=1$.  
 
\ethan{To quantify the error between the reconstructed and LES PSD, we compute an error metric,
\begin{equation}
    \epsilon = \sqrt{\int_{St} \int_{x} \frac{(\mathrm{PSD}_{LES} - \mathrm{PSD}_{Recon})^2}{\mathrm{PSD}_{LES}^2} \mathrm{d}St \mathrm{d}x},
\end{equation}
reflecting the error in acoustic energy. We present this error in figure \ref{fig:NearFieldEnergy} for reconstructions consisting of different numbers of modes. Shown are the errors for two Mach numbers ($M_j=0.9$ and $M_j=1.5$), and three azimuthal wavenumbers, where the filled symbols for the $M_j = 1.5$ case provides a reference between the quantitative measure and the qualitative visualization of figure \ref{fig:M15_Recon_New}. Additionally, in order to assess the utility of the eddy viscosity model, we present results both including the eddy viscosity (filled symbols) and neglecting it (hollow symbols).}  

\ethan{We find significant improvements in reconstructing the near-acoustic field when including the RANS eddy-viscosity field when compared to results using a constant turbulent Reynolds number of $Re_T = 3 \times 10^4$. We note that previous results \citep{pickering2021optimal} only considered RANS eddy-viscosity resolvent models with respect to the dominant near-field hydrodynamic SPOD modes.  While the rank-1 models for the $M_j=1.5$ jet are similar with and without the eddy viscosity, the remaining reconstructions show a strong and clear advantage to the adopted eddy-viscosity approach. Particularly as sub-optimal modes are added to the basis, the eddy-viscosity model converges rapidly toward the LES whereas the turbulent-Reynolds-number model shows little improvement.  This result is consistent with our previous findings \cite{pickering2021optimal}, which showed a more profound effect of the eddy viscosity on sub-optimal modes associated with the Orr-mechanism than on modes associated with the Kelvin-Helmholtz mechanism, where the latter are dominant over most of the frequency-wavenumber space being considered here. }

\ethan{Overall we find the errors related to those with an eddy-viscosity compare favorably when compared to a number of past studies. In particular, the $M_j = 1.5$ case at $m=0$ presents a rank-3 reconstruction with an error of approximately 20\%, while previous studies have reported errors $>30$\% when considering a rank-50 reconstruction for only a single frequency ($St = 0.33$) \cite{jeun2016input}. Our approach and metric considers reconstructions over $St = [0,1]$, rather than one individual frequency.  Comparing the two Mach numbers, it is apparent that a larger number of modes are required to reconstruct the near acoustic field of the $M_j=0.9$ jet. For example, about 10 modes are necessary to obtain a similar quantitative match as compared to just three modes at $M_j=1.5$. This is consistent with multiple past observations where the $M_j = 0.9$ jet possesses non-negligible contributions from suboptimal modes that are correlated, or, as described in the time domain, as being linked via ``jittering'' \citep{cavalieri2011jittering}, thus requiring many modes to reconstruct the acoustic field \citep{freund2009turbulence,towne2015stochastic}.}

\subsection{Far-Field Results} \label{sec:far-field_results}
\begin{figure}
\centering
\vspace{-0.5cm}
	\includegraphics[width=0.495\textwidth,trim={0cm 0cm 0cm 0cm},clip]{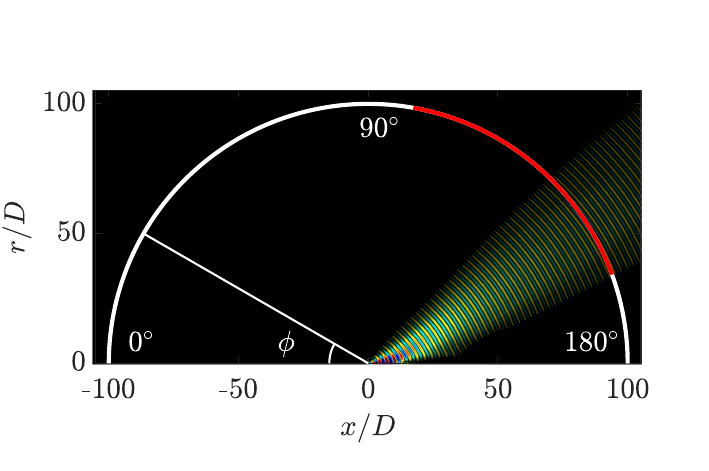}
	\vspace{-0.5cm}
\caption{Schematic of the far-field arc at $100D$ from the nozzle exit. The angle along the arc is defined as $\phi$, with $0^{\circ}$ on the upstream axis and $180^{\circ}$ on the downstream axis. The red portion of the arc denotes the region of interest, $\phi = 100^{\circ}-160^{\circ}$, and the acoustic beam presented is the first resolvent mode for $M_j = 1.5$, $St=0.26$, $m=0$, found for the far-field region.}
\label{fig:Schematic}
\end{figure}

We now extend the eddy-viscosity-enhanced resolvent basis to the far-field, and aim to find the modes that are optimal on an arc $100D$ from the nozzle and a range of polar angles from $\phi=100^{\circ}$ to $\phi = 160^{\circ}$ (where $\phi = 180^{\circ}$ lies on the downstream axis).  The domain is depicted in figure \ref{fig:Schematic} and the details of the associated output matrix, $\mathbf{C}$, are detailed in Appendix \ref{app:far-field}.

Figure \ref{fig:SPOD_Res} presents the magnitude of the first three resolvent modes along the arc for both jets at $St = 0.26$ and $m=0$.  The same three-beam structure apparent in figure \ref{fig:ResolventModes} is evident here, with the dominant one-beam mode peaking at $\phi \approx 150^{\circ}$.  This progression in beam number and location continues in the higher mode numbers not visualized here. Also plotted in figure \ref{fig:SPOD_Res} are the magnitude of modes found via spectral proper orthogonal decomposition (SPOD) of the LES data.  These modes, which optimally reconstruct the CSD of the far-field arc, are useful to compare to the resolvent modes since a close correspondence between resolvent and SPOD modes indicates that the resolvent mode forcings are mutually uncorrelated \citep{towne2018spectral}.  Indeed, we see a reasonable agreement between the far-field SPOD and resolvent modes. The amplitudes and exact locations vary slightly, but such close agreement suggests that an uncorrelated model may suffice.

\begin{figure}
 \vspace{-0.5cm} $M_j = 0.9$ \\  \vspace{-0.35cm} \begin{center} \includegraphics[width=0.475\textwidth,trim={0cm 0cm 0cm 0.2cm},clip]{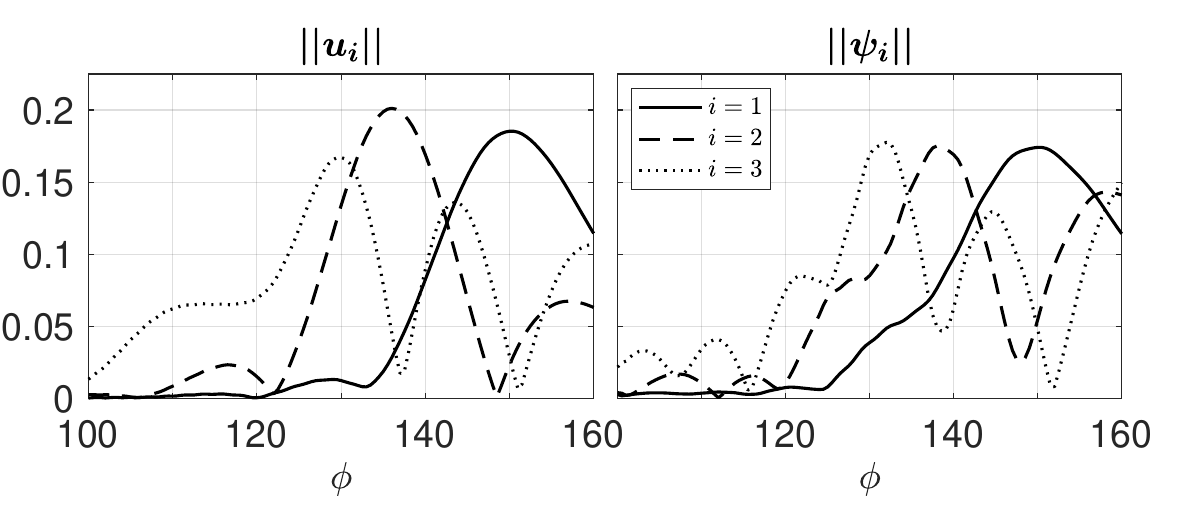} 	\end{center} 
  \vspace{-0.75cm} $M_j = 1.5$ \\  \vspace{-0.35cm} \begin{center} \includegraphics[width=0.475\textwidth,trim={0cm 0cm 0cm 0.2cm},clip]{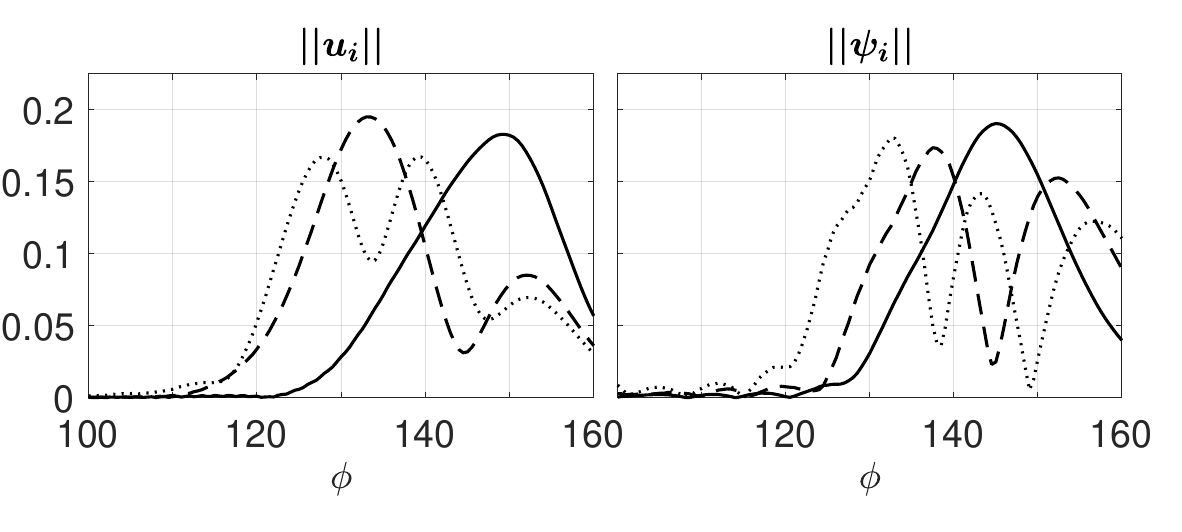} \end{center}  	\vspace{-0.5cm}
\caption{Magnitude of the first three resolvent (left) and SPOD (right) modes computed on the far-field arc for the $M_j=0.9$ (top) and $M_j=1.5$ (bottom) jet at $St=0.26$ and $m=0$.}
\label{fig:SPOD_Res}
\end{figure}

\begin{figure*}
\centering
\vspace{0.5cm}
	\includegraphics[width=1\textwidth,trim={0cm 0cm 0cm 0cm},clip]{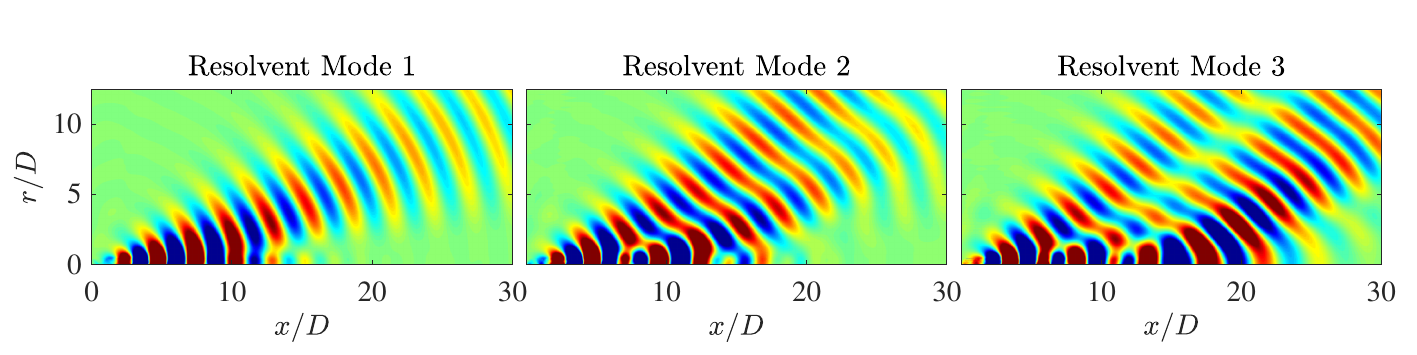}
\caption{The near-field of three resolvent modes of fluctuating pressure, $q_{p'}$, computed considering the 100$D$ arc from $\phi = 100^{\circ}-160^{\circ}$ . Red and blue contours vary from $\pm$ $20\%$ of the maximum fluctuating pressure of each mode, $\pm 0.2 ||q_{p'}||_{\infty}$.  $M_j=1.5$, $St=0.26$, $m=0$.}
\label{fig:ResolventModes_Far}
\end{figure*}

\begin{figure}
\vspace{-0.25cm} \hspace{0.65cm} $M_j = 0.9$ \hspace{2.35cm} $M_j = 1.5$ \\ 
	\includegraphics[width=0.495\textwidth,trim={0cm 0cm 0cm 0cm},clip]{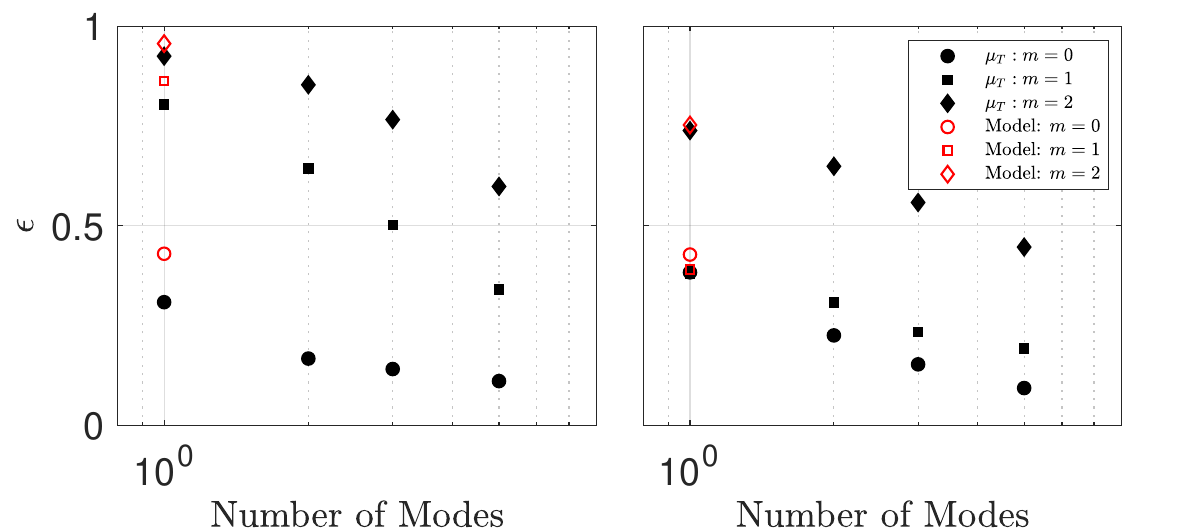}
\caption{\ethan{Error between the LES and reconstructed 100$D$ far-field arc over $\phi = [100,160]$ by number of resolvent modes retained (i.e. $n = [1,2,3,5]$). Filled symbols indicate modes using the RANS eddy-viscosity model, while the red hollow symbols give the error using the rank-1 model.}}
\label{fig:FarField_Error}
\end{figure}

Figure \ref{fig:ResolventModes_Far} shows the near-field signatures of the dominant three far-field modes plotted in figure~\ref{fig:SPOD_Res} for the $M_j=1.5$ jet.  This plot should be directly compared to figure \ref{fig:ResolventModes}, which showed the dominant three near-field modes.  Outside the jet, the modes are nearly indistinguishable.  Within the jet (along the $x$-axis), there are differences that can be associated with the larger hydrodynamic wavepacket imprint left in the near-field modes and missing in the far-field ones, consistent with the empirical results of \citet{towne2015stochastic}.  

We now assess how well the computed resolvent modes reconstruct the PSD of the far-field region across $St \in [0.1, 1]$ and $m = [0,1,2]$ \ethan{in figure \ref{fig:FarField_Error} using the same PSD error metric as figure \ref{fig:NearFieldEnergy}.  For both jets at $m=0$, the rank-1 resolvent reconstruction provides substantial agreement between the LES, at 30\% and 40\% error for $M_j = 0.9$ and 1.5 respectively. The nonzero azimuthal wavenumbers, with the exception of $M_j = 1.5$ and $m=1$, require many additional modes to achieve error levels comparable to the rank-1 $m=0$ error. This higher-rank behavior is similar to what was observed when reconstructing with near-acoustic-field modes for non-zero azimuthal wavenumbers in the previous section.}



\subsection{A simple fit/model} \label{sec:simple}

Considering we may reconstruct (i.e. to 30-40\% error) the far-field acoustics at low-rank, we now ask whether we can define a simple forcing model. One approach would be to propose a form of the forcing cross-spectral density tensor, $\bm{S}_{ff}$,  and project this form onto the resolvent input modes to produce a reduced-order matrix $\tilde{\bm{S}}_{\beta \beta}$. Despite some clear trends for the dependence of $\bm{S}_{ff}$ on mean flow quantities \citep{towne2017statistical}, there does not yet exist a general form for estimating $\bm{S}_{ff}$.  We investigate here an alternative approach of directly estimating $\tilde{\bm{S}}_{\beta \beta}$.  That is, we focus on modeling the expansion coefficients rather than the forcing itself.

The estimated covariance matrix $\tilde{\bm{S}}_{\beta \beta}$ contains the least square reconstruction of the observed data in terms of both the amplitudes and correlations necessary to force each resolvent mode. Where the forcings are uncorrelated, the estimated $\tilde{\bm{S}}_{\beta \beta}$ matrix becomes diagonal and only $n$ coefficients (albeit at each azimuthal wavenumber and frequency) require modeling. However, even if the forcing is uncorrelated, minor errors or discrepancies in the data, data-processing, computation of resolvent mode, etc., result in a full $\tilde{\bm{S}}_{\beta \beta}$ matrix. Further, as rank increases, the statistical uncertainty in the terms becomes greater, reducing our hope for successful modeling. Thus, we explore whether neglecting off-diagonal terms is sufficient for a model, but note that this approach provides no guarantees for success; precisely stated, the approximation is not guaranteed to converge as the number of retained modes is increased \citep{towne2018spectral}.

\begin{figure*}
\centering
 \hspace{0.5cm} $M_j = 0.9$ \hspace{7.25cm} $M_j = 1.5$ \\
   	\includegraphics[width=0.475\textwidth,trim={0cm 0cm 0cm 0cm},clip]{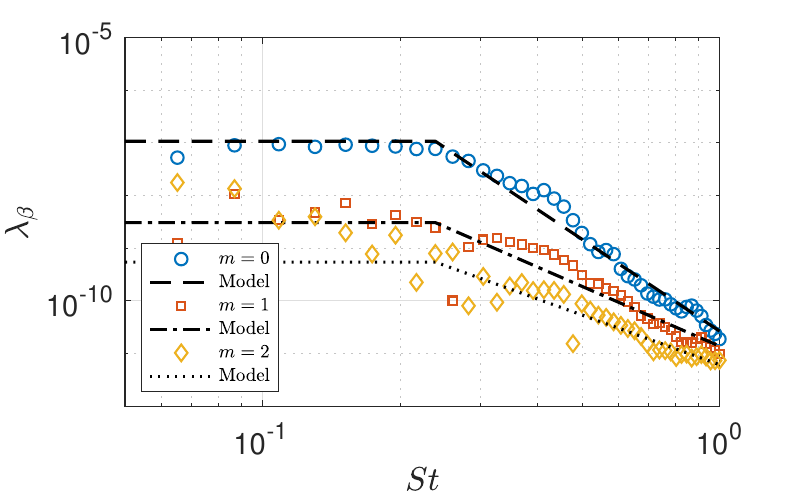}
    \hspace{0.025cm}
	\includegraphics[width=0.475\textwidth,trim={0cm 0cm 0cm 0cm},clip]{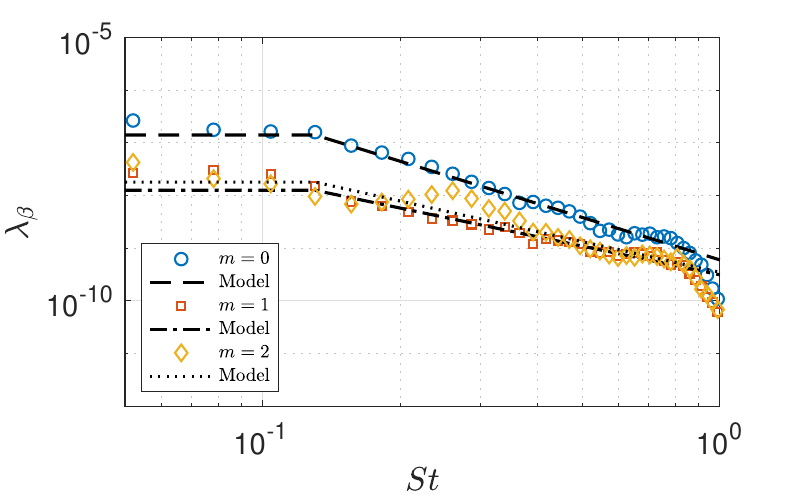}
\caption{Values of the reconstruction projection coefficient, $\lambda_\beta$, of the first resolvent mode for the azimuthal wavenumbers $m=0-2$ and their associated fits, the parameters of which are provided in table \ref{tab:Fits}.}
\label{fig:Spectra_Fits}
\end{figure*}

To limit uncertainty and prevent over-fitting, we assume that the forcing is uncorrelated (i.e. diagonal) and that the projection of the data with the first resolvent mode,
\begin{align}
    \tilde{\bm{S}}_{\beta \beta} &= \bm{\Sigma}_1^{-1} \bm{U}_1^* \bm{S}_{yy} \bm{U}_1 \bm{\Sigma}_1^{-1} = \lambda_\beta,
\end{align}
possesses the lowest uncertainty. These values for the two jets and three azimuthal wavenumbers are shown in figure \ref{fig:Spectra_Fits}. For the $M_j = 1.5$ jet, we see that the forcing amplitudes for $m=0$ and $m=1$ fall upon lines of constant slope for the most acoustically significant frequency ranges, $St = 0.1-0.8$. The $m=2$ data similarly collapse to a line of constant slope, however, the trend is not as clear. Similar observations also hold for the $M_j = 0.9$ jet. We also stress that these curves depend on both the data and the resolvent gains, $\Sigma$. Including the gains is crucial to collapsing the observed trends.

We now look to fit the data with simple curves of the form, 
\begin{align}
    \tilde{\lambda}_{m,\omega} & = a_m St^{b_m}.
\end{align}
For the nonzero azimuthal wavenumbers, the data represent the sum of both the clockwise ($m$) and counter-clockwise ($-m$) directions about the round jet, such that $a_{m}$ represents $a_{+m} + a_{-m}$, or $2 a_{+m}$, since $a_{+m} \approx a_{-m}$, as either rotation about the jet is of equal probability. The exponent, $b_{m}$ is unaffected by this symmetry. Figure \ref{fig:Spectra_Fits} provides the lines of best fit, where the fits are computed over the region of $St = 0.13-0.7$ for the $M_j = 1.5$ and $St=0.22-1$ for the $M_j = 0.9$. \ethan{This choice of low-frequency cutoff is a compromise between keeping the analysis as complete and general as possible, while accounting for the finite computational domain where low-frequency modes are not fully captured within the streamwise extent of the domain, giving the observed plateau in forcing energy. The choice is further justified by noting that the low frequency cutoffs for each case may be related} by adjusting the Strouhal number by Mach number, $St_{0.9} = St_{1.5} \times 1.5/0.9$, meaning each range is associated with the same range of acoustic Strouhal number, $St_{c_\infty}$. The upper bound for the $M_j=0.9$ case extends to $St = 1.17$, however, we cap the upper bound to $St=1$ as done throughout this manuscript. 


Table \ref{tab:Fits} provides the \ethan{fitted} coefficients for each jet and azimuthal wavenumber. At present, we do not have any physical interpretations of these fits, other than the obvious fact that the power law gives an expected decrease in energy as frequency increases (and thus the length scales of the structures decrease). We suspect we may find similar curves via projection of the resolvent forcing modes with the turbulent kinetic energy or other mean-flow quantities, but leave this for future work. 

\begin{table*}
 \hspace{1.1cm} $M_j = 0.9$ \hspace{4.25cm} $M_j = 1.5$ \\
\begin{tabular}{| c | c | c | c | c | c | c |}  
 \hline
 Param. & $m=0$ & $m=1$ & $m=2$ & $m=0$ & $m=1$ & $m=2$ \\  \hline
 $a_m$ & $2.65 \times 10^{-11}$ &  $1.38 \times 10^{-11}$  &  $6.14 \times 10^{-12}$ & $7.10 \times 10^{-10}$ &  $3.89 \times 10^{-10}$  &  $4.66 \times 10^{-10}$  \\ \hline
 $b_m$ & $-5.80$ & $-3.77$ & $-3.13$ & $-2.58$ & $-1.7$ & $-1.76$  \\ \hline  
\end{tabular}
\caption{Fit parameters used for the $M_j = 0.9$ and $M_j = 1.5$ jets shown in figure \ref{fig:Spectra_Fits}.}
    \label{tab:Fits}
\end{table*}

To determine how well such curves predict the data, we use the fitted curves to compute,
\begin{align}
    \tilde{\bm{S}}_{yy}(\phi) = \tilde{\lambda}_{m,\omega} \tilde{\bm{U}} \Sigma^2 \tilde{\bm{U}}^*,
\end{align}
where $\tilde{\bm{U}}$ represents the truncated resolvent basis to rank-$n$. Additionally, as we cannot expect our methods to have accurately captured such large structures in the finite domain used, we use the piece-wise function
\begin{align}
    \tilde{\lambda}_{m,\omega}  &=   a_m St^{b_m}  \hspace{0.65cm}\text{for} \hspace{0.5cm} St > St_{min} \\
    \tilde{\lambda}_{m,\omega}  &=  a_m St_{min}^{b_m} \hspace{0.5cm}   \text{for} \hspace{0.5cm} St \leq St_{min},
\end{align}
where \ethan{$St_{min} = 0.22 $ and $0.13$ for $M_j = 0.9$ and $M_j = 1.5$, respectively. Figure \ref{fig:FarField_Error} provides the reconstruction error for the rank-1 model for both jets and three wavenumbers. For the $M_j = 1.5$ jet, the rank-1 model yields a close approximation of the reconstructions at 40\% for both $m=0$ and $m=1$. For $M_j =0.9$, the $m=0$ error is about 10\% worse than the projections, but still at 40\% for a rank-1 model. The $m=1$ case for the $M_j = 0.9$ jet and both $m=2$ cases perform similarly to the perfect reconstructions, but each of these is relatively poor at reducing the error.} 

\begin{figure*}
 \hspace{0.5cm} $M_j = 0.9$ \hspace{7.25cm} $M_j = 1.5$ \\
\begin{center}
 \vspace{-0.35cm}
   	\includegraphics[width=0.475\textwidth,trim={0cm 0cm 0cm 0.3cm},clip]{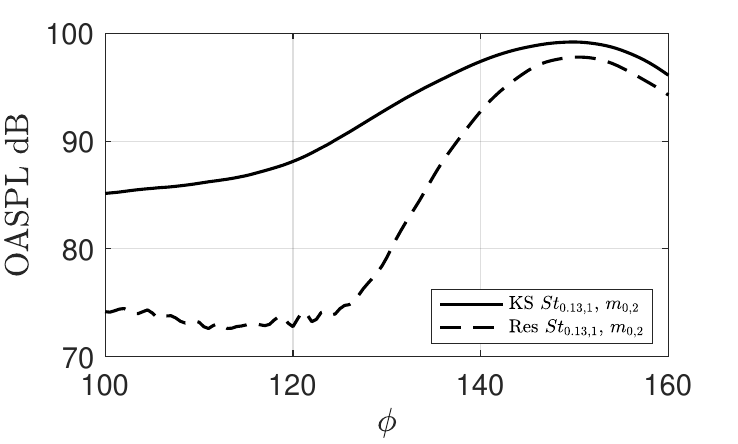}
    \hspace{0.025cm}
	\includegraphics[width=0.475\textwidth,trim={0cm 0cm 0cm 0.3cm},clip]{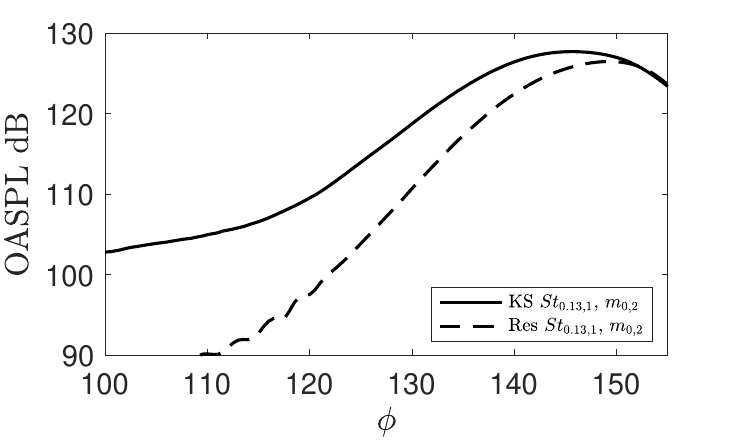}
\end{center}
\caption{OASPL of the $M_j = 0.9$ (left) and $M_j = 1.5$ (right) turbulent jets at $100D$ from the nozzle over the arc $\phi$. The solid black line denotes the \ethan{OASPL from the Kirchhoff surface values found from propagating the near-field LES pressure field, while the solid line gives the resolvent model estimation. In each case, the OASPL only considers the acoustically relevant $m=[0,2]$ and $St = 0.13-1$ contributions.}}
\label{fig:OASPL}
\end{figure*}

\ethan{With the rank-1 prediction in hand, we conclude by computing the overall sound pressure level (OASPL) found by 
\begin{equation}
    \mathrm{OASPL}(\phi) = 10 \text{log}_{10} \bigg(\sum_{St_\mathrm{min}}^{St_\mathrm{max}} 2 \sum_{-m_\mathrm{max}}^{m_\mathrm{max}} \text{diag}(\bm{S}_{yy}(\phi,m,St)) \bigg).
\end{equation}
Figure \ref{fig:OASPL} presents calculations of the OASPL from both jets using the 100$D$ Kirchhoff surface values from the LES and the rank-1 resolvent model considering $m=[0,2]$ and $St = 0.13-1$. For the $M_j = 1.5$ jet, the values are only close at downstream angles and where the jet is the loudest. From $\phi = 140-155$ the model is within 1-3dB. A striking difference is that the resolvent model peaks at an angle of about 5 degrees higher than the LES data. Interestingly, a similar experiment and simulation in \citet{bres2017unstructured} disagreed by the same angle. Although $100D$ data from the experiment is not available, projecting the resolvent modes onto this data would likely result in better alignment, as shifting the LES data by 5 degrees results in a significantly improved estimate (within $0.5dB$ from $\phi = 135-150$). However, as the ultimate source of the discrepancy is unknown, we avoid making any corrections to the model based on these observations.}

\ethan{We see similar behavior between the KS surface and the resolvent model for the transonic case. The rank-1 $m=[0,2]$ resolvent model presents agreement of the peak OASPL to within 2dB at peak noise angles. We stress that this result for the $M_j = 0.9$ jet is rather surprising as many previous studies, although computed in the near-field, found the acoustic field required many modes to agree within 2dB \citep{freund2009turbulence,towne2015stochastic}. This shows that the application of both the KS surface to $100D$ and eddy-viscosity model included in our resolvent analysis significantly reduces the rank of the acoustic jet problem. Further, we note that this transonic jet has been extensively verified by experimental data in the near-field and at $\rho = 50D$, and, although we extend the results to $100D$, the peak angles of the KS and the resolvent model are closely aligned when compared to the $M_j = 1.5$ case.}

\section{Conclusions}

We formulated resolvent analysis to serve as an acoustic analogy by relating the near-field resolvent forcing to both the near- and far-field acoustic regions. Leveraging the availability of an LES database, we examined resolvent-based reconstructions of the acoustic PSD for turbulent $M_j = 0.9$ and $M_j = 1.5$ jets. We represented the forcing cross-spectral density matrix with a truncated set of resolvent modes and approximated the amplitudes of the modes with best-fit expansion coefficients of realizations from the LES acoustic field. We found that models consisting of just a single resolvent mode can accurately reconstruct the acoustic field for the first two azimuthal modes for a $M_j=1.5$ jet and the $m=0$ azimuthal mode for the $M_j=0.9$ jet.  To reconstruct higher azimuthal modes, the resolvent basis must be increased to at least 5 modes (i.e. $m=2$ and $m=1,2$ for $M_j=1.5$ and $M_j = 0.9$, respectively). In both jets, the use of an eddy-viscosity model in the resolvent formulation led to clearly superior results compared to a fixed turbulent Reynolds number. 

Based on the ability of the rank-1 reconstructions to describe the PSD, we investigated a simple model to collapse the forcing coefficients to one scaling function per azimuthal wavenumber (and Mach number). We found that a power law representation, with only an amplitude and an exponent, suffices to model the  coefficient of the optimal resolvent mode. Fortunately, the first resolvent mode contains much of the acoustic energy, and reductions of the gain for this specific mode (related to the KH mechanism) are likely to provide the greatest reductions in the peak noise of the acoustic field. 

The rank-1 $m=[0,2]$ resolvent models estimate the peak noise to within 2dB for both the $M_j = 1.5$ and $M_j = 0.9$ jets \ethan{at peak noise angles}. Further, the ability of the resolvent basis to describe much of the acoustic field with only a handful of modes across multiple Mach numbers, a large range of frequencies, and the acoustically dominant azimuthal wavenumbers is promising. This shows that the resolvent framework already contains the appropriate acoustic functions to describe jet noise. In future work, we will seek a fully predictive model by estimating the forcing coefficients from mean flow quantities available from RANS.

\section*{Acknowledgments}
The authors would like to thank Andr{\'e} Cavalieri, Oliver Schmidt, and Georgios Rigas for many productive discussions on topics related to this paper.  This research was supported by a grant from the Office of Naval Research (grants No. N00014- 16-1-2445 and N00014-20-1-2311) with Dr. Steven Martens as program manager. E.P. was supported by the Department of Defense (DoD) through the National Defense Science \& Engineering Graduate Fellowship (NDSEG) Program. P.J. acknowledges funding from the Clean Sky 2 Joint Undertaking (JU) under the European Union's Horizon 2020 research and innovation programme under grant agreement No 785303. Results reflect only the authors’ views and the JU is not responsible for any use that may be made of the information it contains. The LES study was performed at Cascade Technologies, with support from ONR and NAVAIR SBIR project, under the supervision of Dr. John T. Spyropoulos. The main LES calculations were carried out on DoD HPC systems in ERDC DSRC.

\appendix

\section{Output matrices, $\mathbf{C}$} \label{app:C}
\subsection{Near-field acoustic output matrix} \label{app:near-field}

For analysis of the near-field acoustics, the output matrix $\bm{C}$ is chosen to only include pressure, $p' = \frac{\rho' \bar{T} + \bar{\rho} T'}{\gamma M_j^2}$, in the region $x/D$ = [0, 30], $r/D$ = [5,20]. Ideally, the LES domain would extend from $r/D = [5,20]$ so that the LES could be directly projected onto the resolvent basis; however, the LES database (i.e. the saved data from the LES) only extends to $r/D = 6$. Although one could define an output matrix $\bm{C}$ that only includes that surface at $r/D=6$, the resolvent modes may still contain hydrodynamic behavior (unless allowed to propagate further from the jet), thus we use the larger domain to ensure the modes are entirely acoustic. Using the larger domain presents a clear loss of orthogonality in the space represented by the LES domain, which is alleviated by truncating the modes to $r/D = [5,6]$ (after computing the resolvent SVD) and implementing a Moore-Penrose inverse such that a least-squares fit of the LES in the resolvent basis can be performed.  While previous studies have suggested the use of a filter based on the turbulent kinetic energy of the jet within the input matrix $\bm{B}$ \citep{towne2017statistical}, we take $\bm{B}$ to be identity for both the near- and far-field analyses for the sake of generality.

\subsection{Far-field acoustic output matrix} \label{app:far-field}

To define an input-output relationship from the near-field forcing to the far-field acoustics, we introduce a Kirchhoff surface and apply it as a linear operator. We define three radii: $R$ as the radial coordinate of the near-field cylindrical surface, $r$ as the coordinate pertaining to the far-field cylindrical surface, and $\rho$ representing the distance from the nozzle in spherical coordinates (e.g. $\rho/D = 100$ for this study). As described in \S~\ref{sec:methods_resolvent}, the input-output problem is defined as
\begin{align}
    \bm{q}_{m, \omega} &= \bm{R}_{m, \omega} \bm{B}\bm{f}_{m, \omega}, \\
    \bm{y}_{m, \omega} &= \bm{C}_{R,\rho}\bm{q}_{m, \omega},
\end{align}
where the output matrix $\bm{C}_{R,\rho}$ is the total Kirchhoff operator that maps the near-field cylindrical surface, $R$, to the far-field spherical surface, $\rho$. This operator is linearly composed of many Kirchhoff surfaces, $\bm{C}_{R,r}$, detailed next.

The cylindrical Kirchhoff operator is comprised of several linear operations to ensure accurate results and is defined as,
\begin{align}
    \bm{C}_{R,r} = \bm{D}^* \bm{H}_r\bm{DPTN} \bm{C}_R, 
\end{align}
where $\bm{C}_R$ is a surface selection matrix ($\in \mathbb{R} ^{N_\mathrm{surface} \times 5 N_r N_x}$), $\bm{N}$ is an interpolation matrix from a non-uniform grid to a uniform grid with $\Delta x/D = 0.025$ ($\in \mathbb{R}^{N_\mathrm{uniform} \times N_\mathrm{surface}}$), $\bm{T}$ is a Tukey windowing matrix (using a taper value of 0.75) that extends over the Kirchhoff surface to reduce spectral leakage ($\in \mathbb{R}^{N_\mathrm{uniform} \times N_\mathrm{uniform}}$), $\bm{P}$ is a padding matrix extending the uniform grid with a total of $2^n$ points ($n$ is set to 15) for computing the upstream and downstream wave propagation, as well as ensuring sufficient accuracy in the transform of the initial surface ($\in \mathbb{R}^{2^n \times N_\mathrm{uniform}}$), $\bm{D}$ is the discrete Fourier transform (DFT) matrix ($\in \mathbb{R}^{2^n \times 2^n}$), and $\bm{H}$ contains the derived Hankel functions of the Kirchhoff surface from \citet{freund2001noise}, with entries along the diagonal for each azimuthal wavenumber, for a specified radial distance, $r$, from the surface at $R$ ($\in \mathbb{R}^{2^n \times 2^n}$).

However, the above operator only supports one specified radial distance from the cylindrical surface at $R$, and a linear combination of $\bm{C}_{R,r}$ and a proper selection of streamwise points is required to construct a spherical arc. Thus, the linear expression to construct the total Kirchhoff operator is then
\begin{align}
    \bm{C}_{R,\rho} = \sum_{i=1}^{N_C} \bm{C}_{x_i} \bm{C}_{R,r_i}, 
\end{align}
where $x_i$ represents the streamwise location in the $100 D$ arc and $r_i$ represents the radial extent to which the Kirchhoff surface must propagate from surface $R$ to the far-field arc $\rho$ for the respective streamwise location. Points are defined along the arc from $\phi = 100^{\circ}-160^{\circ}$ with a resolution of $\Delta \phi = 0.5^{\circ}$.

\section{Non-orthogonal projections of resolvent modes} \label{app:non_orthogonal}

The statistical relations presented in \S~\ref{sec:methods_resolvent} are valid when $\bm{U}$ and $\bm{V}$ are orthogonal bases in the same space as $\bm{y}$ and $\bm{f}$, respectively. However, in the case of the near-field calculations, $\bm{U}$ is defined over a larger space than $\bm{y}$ and a pseudo inverse must be constructed to find the least-square solution to the above projections. First, we truncate the output modes $\bm{U}$ to the output space $x/D$ = [0, 30] and $r/D$ = [5,6] in the pressure field and define the associated output matrix as $\bm{C}_{z}$ where $z$ denotes the new restricted space. Applying $\bm{C}_z$ to both the LES data and resolvent modes gives the ensemble of realizations $\bm{z}$ and resolvent modes $\bm{U}_z$. In addition to reducing the domain space, we also truncate the resolvent response basis to a limited set of $n$ modes, as discussed above, represented as $\tilde{\bm{U}}_z$. There are now two important consequences of reducing the resolvent domain from $\bm{C}_y$ to $\bm{C}_z$. The first is a correction to the gain to the domain $\bm{C}_z$. Since both output domains share identical input modes we have,
\begin{align}
  \sigma^{2}_{i,y} &= \frac{\bm{u}^{*}_{i,y} \bm{W}_y \bm{u}_{i,y}}{\bm{v}^{*}_{i,f} \bm{W}_f \bm{v}_{i,f}}, && \sigma^{2}_{i,z} = \frac{\bm{u}^{*}_{i,z} \bm{W}_z \bm{u}_{i,z}}{\bm{v}^{*}_{i,f} \bm{W}_f \bm{v}_{i,f}},
  \end{align}
  and the gain of the new domain is
  \begin{align}
  \sigma^{2}_{i,z} &= \sigma^{2}_{i,y}  \frac{\bm{u}_{i,z} \bm{W}_z \bm{u}^{*}_{i,z}}{\bm{u}^{*}_{i,y} \bm{W}_y \bm{u}_{i,y}}, 
\end{align}
where, by definition, $\bm{u}^{*}_{i,y} \bm{W}_y \bm{u}_{i,y} = 1$. The second is a loss of orthogonality. Fortunately, we may still determine a least squares fit of the data by computing the Moore-Penrose inverse of $\bm{W}_z^{1/2} \tilde{\bm{U}}_z$, $(\bm{W}_z^{1/2} \tilde{\bm{U}}_z)^+ = (\tilde{\bm{U}}_z^*\bm{W}_z \tilde{\bm{U}}_z)^{-1} \tilde{\bm{U}}_z^* \bm{W}_z^{1/2}$, and projecting it onto the CSD of $z$ to estimate  $\tilde{\bm{S}}_{\beta \beta}$ 
\begin{equation}
         \tilde{\bm{S}}_{\beta \beta} = \tilde{\bm{\Sigma}}_z^{-1} (\bm{W}_z^{1/2} \bm{\tilde{U}}_z)^{+*}  \bm{W}_z^{1/2} \bm{S}_{zz}  \bm{W}_z^{1/2} (\bm{W}_z^{1/2} \bm{\tilde{U}}_z)^{+} \tilde{\bm{\Sigma}}_z^{-1}.
         \label{eqn:MP_PI}
\end{equation}
This approach is similar to the one taken by \citet{towne2020resolvent} for assimilating partially observed flow statistics.



\end{document}